\documentclass[12pt]{iopart}

\usepackage{iopams}
\usepackage{setstack}

\newtheorem{prop}{Proposition}

\newcommand{\la}{{\lambda}}
\newcommand{\La}{{\Lambda}}

\newcommand{\eqand}{{\quad \mathrm{and} \quad}}

\newcommand{\del}{{\delta}}
\newcommand{\diag}{{\mathrm{diag}}}

\newcommand{\lb}{{\mathit{\ell}}}
\newcommand{\nb}{{\mathit{n}}}
\newcommand{\mb}[1]{{m_{(#1)}}}
\newcommand{\Phis}{\Phi^\mathrm{S}} 
\newcommand{\Phia}{\Phi^\mathrm{A}} 
\newcommand{\M}[1]{{\overset{#1}{M}}}

\begin{document}
\title{Type II Einstein spacetimes in higher dimensions}
\author{Mark Durkee}
\address{DAMTP, University of Cambridge\\
Centre for Mathematical Sciences,\\
Wilberforce Road, Cambridge, CB3 0WA, UK\\
M.N.Durkee@damtp.cam.ac.uk}


\begin{abstract}
This short note shows that many of the results derived by Pravda \etal (\textit{Class.\ Quant.\ Grav.\ \textbf{24} 4407-4428}) for higher-dimensional Type D Einstein spacetimes can be generalized to all Einstein spacetimes admitting a multiple WAND; the main new result being the extension to include the Type II case.  Examples of Type D Einstein spacetimes admitting non-geodesic multiple WANDs are given in all dimensions greater than 4.
\end{abstract}

\section{Introduction}
In four dimensions, the Petrov classification of spacetimes has for many years been a useful tool for studying exact solutions of the Einstein equations, as well as helping in the construction of new solutions.  A detailed review of this classification and its applications can be found in the textbook \cite{Exact}.

More recently, increased interest in higher-dimensional gravity has led to the development of a higher-dimensional generalization of this classification by Coley \etal \cite{Classification}, based on the classification of components of the Weyl tensor by their \emph{boost weights}.  We refer the reader to the original paper \cite{Classification}, or the review \cite{Coley:Rev} for a more detailed introduction to this formalism, which is valid in all dimensions $\geq 4$.

In this paper we focus on Einstein spacetimes, that is solutions of the vacuum Einstein equations 
\begin{equation}
  R_{ab} = \La g_{ab},
\end{equation}
allowing the possibility of a cosmological constant $\La$.  Algebraic classification has played a useful role in defining and describing higher-dimensional analogues of several known 4d Einstein solutions, for example Robinson-Trautman \cite{RobTraut} and Kundt \cite{Kundt} spacetimes.

The higher-dimensional classification reduces to the standard Petrov classification in 4d, but several important results do not generalize to higher dimensions.

In 4 dimensions, exactly four discrete \emph{principal null directions} (PNDs) exist for any spacetime (that is not conformally flat) and we say that a spacetime is algebraically special iff two or more of them coincide.   The higher-dimensional analogue of a PND is a \emph{Weyl-aligned null direction} (WAND), but WANDs do not always exist, nor are they always discrete.  It is usually said that a higher-dimensional spacetime is \emph{algebraically special} if it admits a WAND.  In this paper we will restrict ourselves to studying Einstein spacetimes admitting a \emph{multiple WAND}, the higher dimensional analogue of a repeated PND.

Furthermore, in $n=4$ dimensions, the Goldberg-Sachs theorem states that for any Einstein spacetime that is not conformally flat, a null vector field is a repeated PND if and only if it is tangent to a shear-free null geodesic congruence.  However, multiple WANDs can be shearing and/or non-geodesic in $n>4$.

Some progress has been made in providing partial generalizations of the Goldberg-Sachs theorem to higher dimensions.  Pravda \etal \cite{Bianchi} proved that the multiple WAND in vacuum Type III and N spacetimes must be geodesic (their argument can be easily extended to all Einstein spacetimes), while Pravda \etal \cite{TypeD} showed that WANDs are geodesic in so-called `generic' Type II or D spacetimes.

The main purpose of this short note is to generalize many of the other results of \cite{TypeD} (that are only known for Type D spacetimes) to any Einstein spacetimes of principal type II.  We find that most of the propositions, and their proofs, generalize in a simple manner; the significance of this result is that it gives us some understanding of the properties of any multiple WAND in arbitrary dimension.  Our results are stated so that they are valid for any Einstein spacetime admitting a multiple WAND.

Additionally, in Section \ref{sec:nongeo} we use some results on product spacetimes to find examples of $(n\geq5)$-dimensional Einstein spacetimes admitting both geodesic and non-geodesic multiple WANDs.  All of these examples turn out to be Type D.

\subsection{Notation and Useful Identities}
Firstly, we very briefly describe the formalism of algebraic classification in higher dimensions, to set up notation and write down some useful identities for reference.  A reader unfamiliar with this formalism is advised to first consult one of \cite{Classification,Coley:Rev, Bianchi}.

We follow the notation of \cite{TypeD}, which is largely similar to that of other papers in the field.  For a $n$-dimensional spacetime, we work in a frame basis of $n$ vectors $\{\lb \equiv \mb{0}, \nb \equiv \mb{1}, \mb{i}\}$ where $\lb$, $\nb$ are null, and $i,j,...=2,...,n-1$ and $\hat{a},\hat{b},...=0,1,2,...,n-1$ are null frame indices transforming under $SO(n-2)$ and $SO(1,n-1)$ respectively.

Any tensor field $T_{ab...c}$ can be decomposed in the frame basis via
\begin{equation}
  T_{\hat{a}\hat{b}...\hat{c}} = T_{ab...c} m_{(\hat{a})}^a m_{(\hat{b})}^b...m_{(\hat{c})}^c.
\end{equation}
This decomposition allows us to classify components of tensors by their \emph{boost weights} under the action of local Lorentz boosts.  We say that a quantity $\phi$ has boost weight $p$ if, under the action of a local Lorentz boost, $\phi \mapsto \la^p \phi$.  Under such a boost, the basis vectors change as
\begin{equation}
  \lb \mapsto \la \lb, \quad \nb \mapsto \la^{-1} \nb, \quad \mb{i} \mapsto \mb{i}
\end{equation}
and hence the boost weight of a null frame component $C_{\hat{a}\hat{b}\hat{c}\hat{d}}$ of the Weyl tensor can be computed by subtracting the number of 1s from the number of 0s in the list $\hat{a}\hat{b}\hat{c}\hat{d}$, so for example $C_{010i}$ has boost weight +1.  The symmetries of the Weyl tensor require that all components have boost weights $\in\{+2,+1,0,-1,-2\}$.

We say that a null vector field $\lb$ is a WAND if the boost weight +2 components $C_{0i0j} = C_{abcd} l^a m_{(i)}^b l^c m_{(j)}^d$ of the Weyl tensor vanish in a null frame containing $\lb$; if the boost weight +1 components also vanish then we say that the WAND is multiple.

A Type II spacetime is one that admits a multiple WAND, with at least some of the boost weight 0 components $C_{01ij}$, $C_{0i1j}$, $C_{0101}$ and $C_{ijkl}$ of the Weyl tensor non-vanishing.  Following \cite{TypeD}, we define the $(n-2)\times(n-2)$ matrix $\Phi_{ij} = C_{0i1j}$, and denote its symmetric and antisymmetric parts by $\Phis_{ij}$ and $\Phia_{ij}$.  The symmetries and tracelessness of the Weyl tensor then imply that
\begin{equation}
  \fl C_{01ij} = 2 C_{0[i|1|j]} = 2\Phia_{ij}, \quad
  C_{0(i|1|j)} = \Phis_{ij} = -\frac{1}{2} C_{ikjk},\quad
  C_{0101} = -\frac{1}{2} C_{ijij} = \Phi_{ii} \equiv \Phi.
\end{equation}

We also define\footnote{Note that this differs from the some other definitions of $\Psi$ in the literature, e.g.\ \cite{Bianchi}, by numerical factors and ordering of indices.} $\Psi_{ijk} = C_{1ijk}$ to describe the boost weight $-1$ components, and note that contracting on the first and third indices gives the other boost weight $-1$ components via
\begin{equation}
  \Psi_j \equiv C_{101j} = C_{1kjk}=\Psi_{kjk}. \label{eqn:psitrace}
\end{equation}
The covariant derivative has null frame components
\begin{equation}
  D \equiv \lb . \nabla, \quad \Delta \equiv \nb . \nabla \eqand \del_i \equiv m_{(i)} . \nabla ,
\end{equation}
while we represent the covariant derivatives of the frame vectors themselves as
\begin{equation}
  L_{ab} \equiv \nabla_b l_a, \quad N_{ab} \equiv \nabla_b n_a \eqand \M{i}_{ab} \equiv \nabla_b m_{(i)a}.
\end{equation}
In analysing the optics of the null vector field $\lb$, it will often be useful to split the $(n-2)\times (n-2)$ matrix $L_{ij}$ into its symmetric and antisymmetric parts, that is to write
\begin{equation}\label{eqn:SAdef}
  L_{ij} = S_{ij}+A_{ij} \eqand S = \tr(S_{ij}) = \tr(L_{ij}).
\end{equation}
The WAND $\lb$ is tangent to a null geodesic congruence iff $L_{i0}=0 \Leftrightarrow Dl = 0$; if this is the case we say that $\lb$ is geodesic.

\section{Main Results}
\subsection{Type II Bianchi Identities}
In any (pseudo-)Riemannian manifold, the Riemann tensor $R_{abcd}$ obeys the differential Bianchi identities $R_{ab[cd;e]} = 0$.  Working in a null frame, the Bianchi identities of a general spacetime in $n$ dimensions are presented in Appendix B of \cite{Bianchi}.

For an Einstein manifold the Weyl tensor is related to the Riemann tensor by
\begin{equation}
   C_{abcd} = R_{abcd} - \frac{\La}{n-1} ( g_{ac}g_{bd}-g_{ad}g_{bc})
\end{equation}
which immediately implies that the Weyl tensor also obeys the differential Bianchi identities, and hence the null frame equations (B.1-B.16) of \cite{Bianchi} apply with $R_{\hat{a}\hat{b}\hat{c}\hat{d}}$ replaced everywhere by $C_{\hat{a}\hat{b}\hat{c}\hat{d}}$.  Pravda et al \cite{TypeD} used these results to prove several general properties about Type D spacetimes; we look to extend many of them to include the Type II case. 

The equations (20), (22) and (27) of \cite{TypeD} are used frequently throughout the paper, so a useful first step is to demonstrate that they also apply in the Type II case.

The equations (B.8, \cite{Bianchi}) and (B.15, \cite{Bianchi}) do not contain any Weyl tensor components of negative boost weight, and hence the equations (20,21, 22) of \cite{TypeD} that are derived from them in the Type D case are immediately valid for general Type II spacetimes.  That is, when terms involving positive boost weight components of the Weyl tensor vanish, (B.8, \cite{Bianchi}) and (B.15, \cite{Bianchi}) imply the algebraic equations
\begin{eqnarray}
  \fl \quad 0 = \Phi_{ij} L_{k0} - \Phi_{ik} L_{j0}+2\Phia_{kj} L_{i0} - C_{isjk} L_{s0}, \label{eqn:D20}\\
  \fl \quad 0 = 2 (\Phia_{jk}L_{im}+\Phia_{mj}L_{ik}+\Phia_{km} L_{ij} +
       \Phi_{ij} A_{mk} + \Phi_{ik} A_{jm} + \Phi_{im} A_{kj} ) \nonumber \\
   +C_{isjk} L_{sm} + C_{ismj} L_{sk} + C_{iskm} L_{sj}\label{eqn:D21}\\
  \fl \quad  0 = S \Phia_{mj} + \Phi A_{jm} - \Phi_{mi} S_{ij} + \Phi_{ji} S_{im}
                    + 2 (\Phia_{im} A_{ij} - \Phia_{ij} A_{im}) + \frac{1}{2} C_{ismj} A_{si}.\label{eqn:D22}
\end{eqnarray}
with (\ref{eqn:D22}) following from contraction of $i$ with $k$ in (\ref{eqn:D21}).
On the other hand, equations (B.5) and (B.12) of \cite{Bianchi} do contain negative boost components, so they are more complicated in the Type II case.  Contracting $m$ and $j$ in (B.12,\cite{Bianchi}) gives
\begin{eqnarray}
\fl  2 D \Phis_{ik} = 4 \Phia_{ij} A_{kj}+\Phi_{kj} L_{ij} + \Phi_{ji} L_{jk} - \Phi_{ki} S - \Phi L_{ik} - 2 \Phis_{is} L_{sk} - 2\Phis_{sk} \M{s} _{i0} - 2\Phis_{is} \M{s}_{k0} \nonumber \\
            +C_{ijks} L_{sj} + \Psi_{jkj} L_{i0} - \Psi_{ikj} L_{j0} + \Psi_{jij} L_{k0}-\Psi_{kij} L_{j0}. \label{eqn:D23}
\end{eqnarray}
Meanwhile, the symmetric part of (B.5, \cite{Bianchi}) simplifies to
\begin{eqnarray}
\fl  2 D \Phis_{ik} = -2\Phi S_{ik} + (-2\Phi_{is}+\Phi_{si}) L_{sk} + (-2 \Phi_{ks}+\Phi_{sk}) L_{si}\nonumber \\
             -2 \Phis_{sk} \M{s}_{i0} - 2 \Phis_{si} \M{s}_{k0}+ \Psi_k L_{i0} + \Psi_i L_{k0} - (\Psi_{kij} + \Psi_{ikj})L_{j0} \label{eqn:D24}
\end{eqnarray}
Taking (\ref{eqn:D23})-(\ref{eqn:D24}) gives
\begin{eqnarray}
 \fl 0 = - \Phi_{ki} S + \Phi L_{ki} + \Phi_{kj} L_{ij} + 4 \Phia_{ij} A_{kj} + (2\Phi_{kj}-\Phi_{jk})L_{ji} + 2\Phia_{ij}L_{jk} \nonumber \\
      +C_{ijks} L_{sj} + (\Psi_{jkj}-\Psi_k)L_{i0} + (\Psi_{jij}-\Psi_i) L_{k0},
\end{eqnarray}
and applying the identity (\ref{eqn:psitrace}) shows that the terms containing $\Psi$ cancel and we recover equation (26,\cite{TypeD}), which is therefore valid for all spacetimes admitting a multiple WAND:
\begin{equation}
  \fl 0 = - \Phi_{ki} S + \Phi L_{ki} + \Phi_{kj} L_{ij} + 4 \Phia_{ij} A_{kj} + (2\Phi_{kj}-\Phi_{jk})L_{ji} + 2\Phia_{ij}L_{jk} + C_{ijks} L_{sj}, \label{eqn:D26}
\end{equation}
Taking the symmetric part recovers (27,\cite{TypeD}), that is we have
\begin{equation}
  \fl 0 = - S \Phis_{ik}+\Phi S_{ik} + \Phis_{ij}S_{jk}+ S_{ij} \Phis_{jk} + 3(\Phia_{ij}S_{jk} - S_{ij}\Phia_{jk}) + C_{ijks}S_{js}. \label{eqn:D27}
\end{equation}

Given these results, we now move on to state generalizations of Propositions 6-11 in \cite{TypeD}.  Note that all of the equations above are valid for any Einstein spacetime admitting a multiple WAND, though most of them are trivially satisfied for Types III, N and O.

\subsection{Geodesity of WANDs}
Pravda \etal \cite{TypeD} used equation (\ref{eqn:D20}) to prove a result about geodesity of multiple WANDs in Type II or Type D Einstein spacetimes.  Here we state the same result in a slightly more general way, noting that when (iii) holds then either (i) holds, or the spacetime is Type III/N in which case all WANDs are geodesic.
\begin{prop}\label{prop:geocond}
  In an Einstein spacetime that is not conformally flat, a multiple WAND $\lb$ is always geodesic if any of the following conditions on boost weight 0 components of the Weyl tensor hold:
  \begin{enumerate}
    \item $\Phia_{ij}$ is non-vanishing.
    \item None of the eigenvalues of $\Phis_{ij}$ are $-\Phi$.
    \item $C_{ijkm}$ vanishes identically.
  \end{enumerate}
\end{prop}

\subsection{Product spacetimes with non-geodesic WANDs}\label{sec:nongeo}
The authors of \cite{TypeD} constructed an example of a Robinson-Trautman Type D spacetime of dimension $n\geq 7$ where one of its WANDs is non-geodesic.

Here, we construct examples of Einstein spacetimes with non-geodesic WANDs in arbitrary dimension $n\geq 5$ using product manifolds.  Godazgar and Reall \cite{Mahdi} discuss a special case of this: non-geodesic WANDs in $dS_3 \times S^{n-3}$.  The notation in this section is that of \cite{TypeD}.

Let $(M_1,g_1)$ be an ($r\geq3$)-dimensional Lorentzian Einstein spacetime, satisfying
\begin{equation}
  R_{(1)AB} = \La g_{(1)AB}, \quad \quad A,B,...=0,1,r-1 \label{eqn:ein1}
\end{equation}
and $(M_2,g_2)$ be an $(n-r)$-dimensional Riemannian Einstein space, with 
\begin{equation}
  R_{(2)IJ} = \La g_{(2)IJ}, \quad \quad I,J,...=r,...,n-1. \label{eqn:ein2}
\end{equation}

Now, we can construct an $n$-dimensional Einstein spacetime $M=M_1 \times M_2$ with block-diagonal metric $g = g_{(1)} + g_{(2)}$.  Let $\lb$ and $\nb$ be any smooth null vector fields on $M_1$ satisfying $\lb.\nb=1$.  These can be trivially extended to vector fields on $M$ that are everywhere orthogonal to $M_2$ and dependent only on coordinates $x^A$.  Clearly there are examples of such vector fields that are non-geodesic.

Using the equations (12), (14) of \cite{TypeD} for components of the Weyl tensor in a frame basis containing $\lb$,$\nb$, and applying the Einstein equations (\ref{eqn:ein1}) and (\ref{eqn:ein2}), we find that the only non-vanishing components with positive boost weight are
\begin{equation}
  C_{0\hat{A}0\hat{B}} = C_{(1)0\hat{A}0\hat{B}}, \quad C_{010\hat{A}} = C_{(1)010\hat{A}} 
                         \eqand C_{0\hat{A}\hat{B}\hat{C}} = C_{(1)0\hat{A}\hat{B}\hat{C}},
\end{equation}
where $\hat{A},\hat{B},...=2,...,r-1$, while the non-vanishing components with negative boost weight are
\begin{equation}
    C_{1\hat{A}1\hat{B}} = C_{(1)1\hat{A}1\hat{B}}, \quad C_{101\hat{A}} = C_{(1)101\hat{A}} 
                         \eqand C_{1\hat{A}\hat{B}\hat{C}} = C_{(1)1\hat{A}\hat{B}\hat{C}}.
\end{equation}
Therefore, if $M_1$ is conformally flat, then all Weyl tensor components of $M$ with non-zero boost weight vanish in a frame basis associated with \emph{any} null vector fields $\lb$ and $\nb$ tangent to $M_1$.  So, if we choose $\lb$ and $\nb$ to be non-geodesic, we have a pair of non-geodesic multiple WANDs for $M$.

Studying the boost weight 0 components of $C_{abcd}$, we find that $M$ is never conformally flat if $\La\neq 0$.  Therefore, for $\La\neq0$ we can find a large class of simple examples of Type D spacetimes admitting both geodesic and non-geodesic WANDs.  Note that these WANDs are never discrete, since the property of being a WAND is always preserved under null rotations within $M_1$.

More explicitly we have, at least locally,
\begin{equation}
  M = \cases{dS_r \times M_+, \quad \La >0,\\ AdS_r \times M_-, \quad \La <0}
\end{equation}
where $M_\pm$ are $n-r$ dimensional Einstein spaces with positive or negative cosmological constant respectively.  Note that all of the analysis of this section is purely local, so although $M_1$ must be locally isometric to $(A)dS_r$, Einstein spacetimes with more interesting global structures are allowed.

Note that when $M_\pm$ has dimension 2 or 3, the Weyl tensor $C_{(2)}$ also vanishes, and $M_\pm$ is a space of constant curvature, that is it is locally isometric to $S^{n-r}$ or $H^{n-r}$.  Therefore, in $n=5$ dimensions, the only examples of spacetimes of this form are $dS_3 \times S^2$ and $AdS_3 \times H^2$, while for $n=6$, there is $dS_3\times S^3$, $dS_4\times S^2$, $AdS_3\times H^3$ and $AdS_4\times H^2$.  For $n\geq 7$ there are many more possibilities.

When $\La=0$, $M$ is conformally flat iff $C_{(2)\hat{I}\hat{J}\hat{K}\hat{L}} =0$.  Therefore, $M_2$ must have dimension 4 or more to give non-trivial examples in this case, and hence we must have $n\geq 7$.  In this case, this gives many examples of vacuum spacetimes admitting non-geodesic multiple WANDs, namely any spacetime of the form
\begin{equation}
  M = \mathrm{Mink}_r \times M_2, \qquad r\geq 3,\quad \mathrm{dim}(M_2) = n-r\geq 4 
\end{equation}
with $M_2$ any non-flat Einstein space with Riemannian signature.

\subsection{Shearfree, expanding multiple WANDs}\label{sec:shearfree}
When the multiple WAND $\lb$ is shearfree and expanding, further progress can be made in constraining its properties:
\begin{prop}\label{prop:shearfree}
  In an Einstein spacetime of arbitrary dimension $n\geq4$, a shearfree and expanding multiple WAND is geodesic whenever $\Phi_{ij} \neq 0$, and the symmetric part of $\Phi_{ij}$ must take the form
  \begin{equation}\label{eqn:phitrace}
    \Phis_{ij} = \frac{\Phi}{n-2} \del_{ij}.
  \end{equation}
\end{prop}
\noindent\emph{Proof:} Inserting
\begin{equation}
  S_{ij} = \frac{S}{n-2} \del_{ij} \neq 0
\end{equation}
into \eref{eqn:D27} gives the form \eref{eqn:phitrace}.  By Proposition \ref{prop:geocond} this implies that $\lb$ is geodesic  (c.\ f.\ Proposition 7 of \cite{TypeD}). $\Box$

Therefore, \emph{non-geodesic} shearfree and expanding multiple WANDs must have $C_{ijkl}\neq 0$.  The second proposition of \cite{TypeD} uses the Bianchi equation (B.3,\cite{Bianchi}), which is not valid in general Type II spacetimes.  The end result however does generalize, that is:
\begin{prop}\label{prop:imps}
  In an Einstein spacetime of dimension $n>4$, the following implications hold for shearfree and expanding multiple WANDs:
  \begin{enumerate}
    \item $A_{ij} = 0 \Rightarrow \Phia_{ij} = 0$.
    \item $\Phia_{ij}=0, \Phis_{ij} \neq 0 \Rightarrow A_{ij}=0$.
    \item $\Phia_{ij}=0, \Phis_{ij} \neq 0 \Rightarrow C_{ijkm} A_{km}=0$.
  \end{enumerate}
\end{prop}
\noindent \emph{Proof:} To prove (i), (iii) we can apply precisely the same argument as \cite{TypeD}, relying only on the Bianchi equation (\ref{eqn:D22}) in the shearfree case.  For (ii), note that when positive boost weight Weyl tensor components vanish, (B.3,\cite{Bianchi}) reads
\begin{equation}
\fl \quad -2D\Phia_{ij}=2\Phi A_{ij}+4\Phia_{[i|s} L_{s|j]} + 2\Phi_{[i|s} L_{s|j]}+2\Psi_{[i}L_{j]0} + \Psi_{sij} L_{s0} + 4\Phia_{[i|s} \M{s}_{|j]0}. \label{eqn:B3}
\end{equation}
In the case $\Phis_{ij}\neq 0$ Proposition \ref{prop:shearfree} tells us that $\lb$ is geodesic, and hence $L_{i0}=0$.  Putting this, as well as the shearfree condition and $\Phia_{ij}=0$, into (\ref{eqn:B3}) gives
\begin{equation}
  2\Phi A_{ij} + 2\Phis_{[i|s} L_{s|j]}=0 \Rightarrow 2\Phi \frac{n-1}{n-2} A_{ij} = 0
\end{equation}
and hence (ii) holds, which completes the proof.$\Box$

\section{Results in 5 dimensions}
In any 5-dimensional spacetime, the matrix $\Phi_{ij}$ is sufficient to describe all boost weight zero components of the Weyl tensor; the components $C_{ijkl}$ are determined explicitly \cite{TypeD} in terms of $\Phi_{ij}$ by
\begin{equation}
  C_{ijkl} = 2(\del_{il}\Phis_{jk}-\del_{ik}\Phis_{jl}-\del_{jl}\Phis_{ik}+\del_{jk}\Phis_{il})
                    - \Phi (\del_{il}\del_{jk} - \del_{ik} \del_{jl}). \label{eqn:5d}
\end{equation}

This tells us that in $n=5$, for an Einstein spacetime admitting a multiple WAND, either $\Phi_{ij}\neq 0$ or the spacetime is Type III, N or conformally flat (Type O).  Therefore, we have an immediate 5-dimensional corollary of Proposition \ref{prop:shearfree} above:
\begin{prop}
  In a 5-dimensional Einstein spacetime that is not conformally flat, any shearfree and expanding multiple WAND is geodesic.
\end{prop}

Now, for the rest of the section we will make use of the following equations, valid only in $n=5$, obtained from (\ref{eqn:D20}), (\ref{eqn:D22}) and (\ref{eqn:D27}) via insertion of equation (\ref{eqn:5d}), giving
\begin{eqnarray}
\quad \quad \quad (\Phis_{ij} - \Phi \del_{ij} ) L_{k0} L_{k0}  + 2\Phi L_{i0} L_{j0}  &=& -\frac{1}{3} \Phia_{ij} L_{k0} L_{k0} \label{eqn:D20a}\\
\fl \Phia_{ij}(S_{jk}+2A_{jk}) +\Phia_{jk}(S_{ji}+2A_{ji}) \nonumber \\
\fl \quad \quad \quad \quad \quad  +\Phis_{ij}(S_{jk}-2A_{jk})+\Phis_{kj}(-S_{ji}+2A_{ji})&=&\Phia_{ik} S + 2\Phi A_{ki}\label{eqn:D22a}\\
\quad \quad \quad \quad 3(\Phi_{ij}S_{jk} + \Phi_{kj}S_{ji}-S \Phis_{ik}) &=& \del_{ik}(2S_{js}\Phis_{js}-S\Phi). \label{eqn:D27a}
\end{eqnarray}
respectively, these are analagous to (44)-(47) of \cite{TypeD}.\footnote{Note that the rhs of (\ref{eqn:D20a}) is missing in \cite{TypeD}, though this has no effect on the results that follow.}

\subsection{Non-geodesic WANDs in 5d}
Given equation (\ref{eqn:5d}), it is natural to ask whether we can make any further progress in restricting the non-geodesic case in 5d.  The following two propositions are generalizations of those given for Type D spacetimes in \cite{TypeD}.
\begin{prop}\label{prop:nongeo5}
  A 5-dimensional, not conformally flat, Einstein spacetime admitting a non-geodesic multiple WAND $\lb$ has $\Phia_{ij} = 0$, and $\Phis_{ij}$ has matrix eigenvalues $\Phi$, $\Phi$ and $-\Phi$.  In a basis where $L_{i0} = \del_{i4}L_{40}$, we have $\Phis_{ij} = \diag(\Phi,\Phi,-\Phi)$ and the optics of $\lb$ are described by
  \begin{equation}
    L_{ij} = \left( \begin{array}{ccc}
                      0 & 0 & 0\\
                      0 & 0 & 0\\
                      L_{42} & L_{43} & L_{44}
                    \end{array}\right)
  \end{equation}
  for some $L_{42}$, $L_{43}$ and $L_{44}$.
\end{prop}
{\noindent \it Proof:} By Proposition \ref{prop:geocond}, we have $\Phia_{ij} = 0$ whenever $\lb$ is non-geodesic.  Applying an $SO(n-2)$ rotation to the spacelike frame vectors $\mb{i}$, we can move to a basis where $L_{i0}$ is tangent to one of the basis vectors (say $\mb{4}$).  In this new basis, $L_{20}=L_{30}=0$ and $L_{40}\neq 0$.  Using this, equation \eref{eqn:D20a} becomes
\begin{equation}
  (\Phis_{ij}-\Phi\del_{ij} + 2\Phi \del_{i2}\del_{j2})L_{20}L_{20} = 0.
\end{equation}
This implies that 
\begin{equation}\label{eqn:nongeophi}
  \Phis_{ij} = \diag(\Phi,\Phi,-\Phi)
\end{equation}
in this basis, and hence fixes the eigenvalues.

Now we can insert \eref{eqn:nongeophi} into \eref{eqn:D22a} and \eref{eqn:D27a} to fix the form of $L_{ij}$.  In particular, the $44$ component of \eref{eqn:D27a} gives $S=S_{44}$ (recall \eref{eqn:SAdef}), and then the $22$, $23$ and $24$ components give $S_{22} = S_{23} = S_{33} = 0$.

The $23$ component of \eref{eqn:D22a} gives $A_{23}=0$, and then the $24$ and $34$ components show that $A_{24} + S_{24} = 0 = A_{34} + S_{34}$, which completes the proof.$\Box$

\subsection{Other results in 5 dimensions}
The results in Sections 6.2 and 6.3 of \cite{TypeD} also generalize to all 5d spacetimes admitting a multiple WAND, whether geodesic or not, as follows.  In all expressions in this section, indices in brackets (e.g.\ $s_{(i)}$) are excluded from the Einstein summation convention.

\begin{prop} \label{prop:nontwist}
  Let $\lb$ be a non-twisting ($A_{ij}=0$) multiple WAND in a 5-dimensional spacetime, with the expansion matrix $S_{ij}$ having eigenvalues $s_{(i)}$ and trace $S$.  Then the following implications hold:
  \begin{enumerate}
    \item $ s_{(i)} \neq S/2$ for all $i \Rightarrow \Phia_{ij}=0$ and $S_{ij}$, $\Phis_{ij}$ can be simultaneously diagonalized.
    \item $ s_{(4)} = S/2 $, $s_{(2)}, s_{(3)}\neq S/2$ $\Rightarrow$  $\Phi_{ij}= \Phi \del_{4i} \del_{4j}$
    \item $s_{(2)}=s_{(3)}=S/2$, $s_{(4)}=0$ $\Rightarrow$ $\Phis_{i4}=\Phia_{i4}=0$ for all $i$.
\end{enumerate}
\end{prop}
Note that the ordering of the components 234 in the above is of course arbitrary.\\
\noindent \emph{Proof:} The proof of \cite{TypeD}, Proposition 10 (and the special cases below) relies only on equations (\ref{eqn:D22a}) and (\ref{eqn:D27a}), which we have derived again in the Type II case. Therefore, the results there generalize directly.$\Box$

Finally, \cite{TypeD} proves a result that is a partial converse of this proposition, looking to categorize Type D Einstein spacetimes with $\Phia_{ij}=0$.  The proof of that relies on the Bianchi equation (B.3,\cite{Bianchi}) which contains components of negative boost weight, and hence its generalization to all spacetimes admitting a multiple WAND requires a little extra thought, although most parts of it are largely similar.  Working in a frame $\Phis_{ij}=\diag(p_{(2)},p_{(3)},p_{(4)})$, equations (\ref{eqn:D22a}), (B.3,\cite{Bianchi}) and (\ref{eqn:D27a}) reduce to
\begin{eqnarray}
  S_{ik} (p_{(i)}-p_{(k)}) + A_{ik}(-2p_{(i)}-2p_{(k)}+2\Phi) = 0,\label{eqn:D61}\\
  S_{ik} (p_{(i)}-p_{(k)}) + A_{ik}(p_{(i)}+p_{(k)}+2\Phi) = 2\Psi_{[i}L_{k]0} + \Psi_{sik}L_{s0},\label{eqn:D62}\\
  S_{ik} (p_{(i)}+p_{(k)}) = \frac{1}{3} \del_{ik} (3S p_{(i)}+2\Phis_{jm}S_{jm} - S\Phi)\label{eqn:D63}
\end{eqnarray}
respectively.  We can use these equations to prove:
\begin{prop}
  A 5-dimensional Einstein spacetime that is not conformally flat, admits a multiple WAND, and has $\Phia_{ij}=0$, meets the conditions for at least one of the cases $(a)-(g)$ of Table \ref{tab:prop11}.  The multiple WAND then has geodesity and optical properties described by Table \ref{tab:prop11}.
\end{prop}
\noindent \emph{Proof:}
In case (a), we know from Proposition \ref{prop:nongeo5} that $\lb$ must be geodesic.  Therefore, the rhs of \eref{eqn:D62} vanishes, and subtracting \eref{eqn:D61} from \eref{eqn:D62} implies that $A_{ik}=0$.

Equation \eref{eqn:D63} gives
\begin{equation}
  S_{ik} = \del_{ik} \left[\frac{S}{2} + \frac{2\Phis_{jm}S_{jm} - S\Phi}{6 p_{(i)}} \right] \equiv s_{(i)} \del_{ik}
\end{equation}
and hence $S_{ik}$ and $\Phis_{ik}$ can be simultaneously diagonalized.  This leaves us with 3 simultaneous equations
\begin{equation}
  \fl 6 \, p_{(i)}\left(s_{(i)} - \frac{s_{(2)} + s_{(3)} + s_{(4)}}{2}\right) = (2 p_{(2)}-\Phi) s_{(2)}+(2 p_{(3)}-\Phi) s_{(3)}+(2 p_{(4)}-\Phi) s_{(4)}
\end{equation}
for the 3 eigenvalues $s_{(i)}$, which can be solved simultaneously to fix the eigenvalues in terms of the trace $S$, as:
\begin{equation}
  s_{(i)} = \frac{p_{(i)}(\Phi-p_{(i)})S}{2(p_{(2)}p_{(3)}+p_{(3)}p_{(4)}+p_{(4)}p_{(2)})}.
  \label{eqn:casea}
\end{equation}
The other cases can be analysed in a similar way.  Note that the analysis only differs from that in \cite{TypeD} when the rhs of \eref{eqn:D62} is non-vanishing, and by Proposition \ref{prop:nongeo5} this can happen only in case (e), in which case the analysis was covered in the proof of that proposition.$\Box$
\begin{table}[htb]
  \caption{\label{tab:prop11}Restrictions on the geodesity and optics of the multiple WAND $\lb$ when $\Phia_{ik}=0$.  All matrices are given in the frame where $\Phis_{ik} = \diag(p_{(2)},p_{(3)},p_{(4)})$.}
  \begin{tabular}{cl|cll}
  \br
  Case & Conditions & Geo? & Expansion/Shear & Twist\\
  \mr
  (a) & $\Phi_{ij} = \diag(p_{(2)},p_{(3)},p_{(4)})$ & Yes & $S_{ij}= \diag(s_{(2)},s_{(3)},s_{(4)})$ & $A_{ij}=0$\\
    & with $p_{(i)}+p_{(j)} \neq 0 \; \forall i,j$ & & (see equation (\ref{eqn:casea})) & \\
  (b) & $\Phi_{ij} = \diag(0,p,\Phi-p)$ & Yes & $S_{ij}=\diag(0,S/2,S/2)$ & $A_{ij}=0$\\
    & with $\Phi \neq 0$, $p\neq 0$  &     &  & \\
  (c) & $\Phi_{ij}=\diag(0,p,-p)$ & Yes & $S_{ij}=\diag(0,S/2,S/2)$ & $A_{23}=A_{24}=0$\\
    & with $p \neq 0$           &     &                             & \\
  (d) & $\Phi_{ij} = \diag(\Phi,p,-p)$  & Yes & $S_{23} = S_{24} = S_{22}= 0$  & $A_{23}=A_{24}=0$,\\
    & with $p \neq \pm \Phi,0$ & & $S_{33} = \frac{1}{2}S\left(1-\Phi/p\right)$, & $\Phi A_{34} = -p S_{34}$ \\ 
    &                       &      & $S_{44} = \frac{1}{2}S\left(1+\Phi/p\right)$ & \\
  (e) & $\Phi_{ij}=\diag(\Phi,\Phi,-\Phi)$ & Yes/No & $S_{22}=S_{33}=S_{23}=0$   & $A_{23}=0$\\
    & with $\Phi\neq 0$            &     & others arbitrary                    & $A_{24}+S_{24}=0$\\
    &                              &     & (cf Proposition \ref{prop:nongeo5}) & $A_{34}+S_{34}=0$\\
  (f) & $\Phi_{ij}=\diag(\Phi,0,0)$  & Yes & $S_{23}=S_{24}=0$                   & $A_{ij}=0$\\
    &                              &     & $S_{22}=S/2$, others arb.           & \\ \cline{4-5}
  (g) & $\Phi_{ij}=0$                & Yes & (Type III/N, see \cite{Bianchi})     & \\
  \br
  \end{tabular}
\end{table}\begin{Large}\begin{Large}                        \end{Large}           \end{Large}

\newpage
\section*{Acknowledgements}
I would like to thank Harvey Reall and Mahdi Godazgar for useful discussions, as well as Vojt\v ech Pravda and Marcello Ortaggio for helpful comments on a draft version.  I am supported by the Science and Technology Facilities Council.

\end{document}